\documentclass[amsmath,amssymb,aps,pra,superscriptaddress,showpacs,10pt,reprint,longbibliography]{revtex4-1}

\usepackage[table]{xcolor}
\usepackage{booktabs}

\usepackage{graphicx}

\usepackage[colorlinks,linkcolor=blue,citecolor=blue,urlcolor=blue]{hyperref}

\def\bg#1\eg{\begin{align}#1\end{align}}

\newcommand{\I}{i}

\newcommand{\bra}[1]{\left\langle{#1}\right|}
\newcommand{\ket}[1]{\left|{#1}\right\rangle}
\newcommand{\braket}[2]{\langle{#1}|{#2}\rangle}

\begin{document}

\title[]{Phase sensitivity of gain-unbalanced nonlinear interferometers}

    \collaboration{Published in
    \href{https://link.aps.org/doi/10.1103/PhysRevA.96.053863}
    {Physical Review A \textbf{96}, 053863  [2017]}.}

\author{Enno Giese}
\affiliation{Department of Physics, University of Ottawa, 25 Templeton Street, Ottawa, Ontario K1N 6N5, Canada.}
\author{Samuel Lemieux}
\affiliation{Department of Physics, University of Ottawa, 25 Templeton Street, Ottawa, Ontario K1N 6N5, Canada.}
\author{Mathieu Manceau}
\affiliation{Max-Planck-Institute for the Science of Light, Staudtstra{\ss}e 2, 91058 Erlangen, Germany.}
\author{Robert Fickler}
\affiliation{Department of Physics, University of Ottawa, 25 Templeton Street, Ottawa, Ontario K1N 6N5, Canada.}
\author{Robert W. Boyd}
\affiliation{Department of Physics, University of Ottawa, 25 Templeton Street, Ottawa, Ontario K1N 6N5, Canada.}
\affiliation{Institute of Optics, University of Rochester, Rochester, New York 14627, USA.}

\begin{abstract}
The phase uncertainty of an unseeded nonlinear interferometer, where the output of one nonlinear crystal is transmitted to the input of a second crystal that analyzes it, is commonly said to be below the shot-noise level but highly dependent on detection and internal loss.
Unbalancing the gains of the first (source) and second (analyzer) crystals leads to a configuration that is tolerant against detection loss.
However, in terms of sensitivity, there is no advantage in choosing a stronger analyzer over a stronger source, and hence the comparison to a shot-noise level is not straightforward.
Internal loss breaks this symmetry and shows that it is crucial whether the source or analyzer is dominating.
Based on these results, claiming a Heisenberg scaling of the sensitivity is more subtle than in a balanced setup. 
\end{abstract}

\maketitle

\date{\today}

\section{Introduction}

A nonlinear interferometer (NLI)---characterized by the Lie group SU(1,1) and consisting of two consecutive nonlinear crystals~\cite{Yurke86}---is a potential alternative to a linear interferometer seeded by a squeezed state~\cite{Caves81} for high-precision measurements because of its supreme phase sensitivity~\cite{Chekhova16}.
Indeed, it is said to feature a `Heisenberg scaling' when the gains in both crystals are equal~\cite{Yurke86,Linnemann16}.
It has been suggested that one can suppress the influence of detection loss by using unbalanced gains in the two crystals~\cite{Sparaciari16,Manceau17}, and in fact sub-shot-noise phase sensitivity in an unseeded NLI with this method was recently demonstrated~\cite{Manceau17PRL}. 

In this article, we show that (i) the photon statistics of an unseeded and gain-unbalanced NLI lead to a suppression of the deleterious influence of detection loss, (ii) the phase sensitivity is in this case ultimately limited by the \emph{lower} gain and is therefore symmetric with respect to the two crystals, (iii) a comparison to the shot-noise level is not straightforward, and (iv) internal loss breaks the symmetry so that a \emph{higher} gain in the source crystal might be beneficial.
Since NLIs may be intrinsically gain-unbalanced, claiming a Heisenberg scaling has to be carefully justified in each individual case.

An NLI characterized by the SU(1,1) group typically consists of two nonlinear crystals A and B, as shown in Figs.~\ref{fig_degenerate} and~\ref{fig_nondegenerate}.
In the original proposal~\cite{Yurke86}, crystal A is the \emph{source} of the radiation, which is transmitted into crystal B acting as an \emph{analyzer}.
The NLI can be operated at constructive interference where both crystals generate radiation, a method that has, for example, been explored to create and tailor bright squeezed vacuum states of light~\cite{
Perez14,Sharapova15,Lemieux16,Beltran17}.
In addition, it was shown that by seeding the NLI with a light field, the phase sensitivity is boosted even further---for both a coherent- and a squeezed-state input field~\cite{Plick10,Li14,Sparaciari16}---and the influence of internal loss may be decreased~\cite{Marino12,Ou12}.
To focus on the physical mechanisms of an NLI, we restrict ourselves in this article to the unseeded case with vacuum input modes, which has no correspondence in a conventional interferometer.

As it is the case in other quantum physical processes with multiple nonlinear crystals, such as induced coherence~\cite{Wang91,Kolobov17}, it is essential that the two crystals are pumped coherently.
Nevertheless, the gains in both crystals can be controlled separately and the relative phase of the pump field can be varied.
In fact, it would be experimentally difficult to ensure that the gain of the source and the gain of the analyzer are \emph{exactly equal}, especially since the number of photons produced scales exponentially with the electric field amplitude of the pump.
Unbalanced gains give an additional degree of freedom to optimize the properties of the NLI.
In this spirit, it was shown theoretically that the deleterious effects of detection loss~\cite{Marino12} can be overcome by intentionally \emph{unbalancing} the gains~\cite{Sparaciari16,Manceau17,Anderson17b}.
This effect was recently demonstrated experimentally~\cite{Manceau17PRL}, for the case in which the analyzer is pumped more strongly than the source.
On the other hand, the significance of the analyzing crystal is questioned by proposals to operate the device in a truncated mode of operation---with only the source as a squeezer~\cite{Anderson17}.
At first sight, these considerations imply an opposite role of the analyzing crystal and make it necessary to investigate the effect of gain unbalancing in more detail to understand the ultimate limit of the sensitivity of the device.
A seeded and gain-unbalanced setup has been investigated in~\cite{Kong13}, but without explicit consideration of the limitations on the sensitivity.

In Sec.~\ref{sec_Theoretical_description} we use simple transformations to derive exact analytical expressions for the detected photon number, its variance, and the phase sensitivity that can be applied to a situation with unbalanced gains, and we show the conditions under which detection loss is significant or can be overcome.
In Sec.~\ref{sec_Phase_sensitivity_for_unbalanced_gain} we calculate the phase sensitivity of a gain-unbalanced NLI.
The effect of internal loss breaks the symmetry between source and analyzer so that---depending on the parameters of the setup---it makes a difference which crystal is pumped more strongly, as we show in Sec.~\ref{sec_Breaking_the_symmetry}.
Because all of these calculations focus on a degenerate NLI, we generalize in Sec.~\ref{sec_Comparison} our approach to compare the results to a nondegenerate setup, before we conclude in Sec.~\ref{sec_Conclusions}.
To keep this paper self-contained, we include the detailed calculations for the degenerate NLI in appendix~\ref{app_Degenerate_configuration}, the quantum Fisher information in a lossless and balanced setup in appendix~\ref{app_Quantum_Fisher_information}, and the nondegenerate NLI in appendix~\ref{app_Non-degenterate_configuration}.

\section{Theoretical description}
\label{sec_Theoretical_description}

\begin{figure}[htb] \centering
\includegraphics[scale=1]{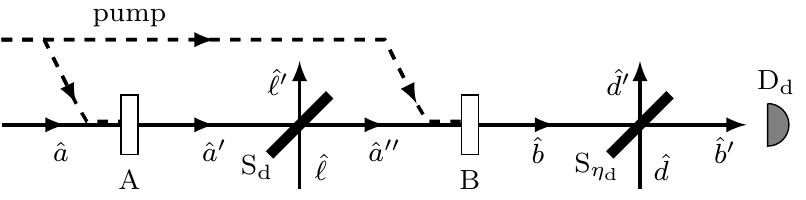}
    \caption{
    Schematic of a degenerate nonlinear interferometer that consists of two coherently pumped nonlinear crystals A and B.
    Internal loss is modeled by a beam splitter S$_\text{d}$, detector inefficiencies by a beam splitter S$_{\eta_\text{d}}$.
    The bosonic annihilation operator $\hat{a}$ denotes the input mode of the interferometer, the operator $\hat{b}'$ the field detected by detector D$_\text{d}$.
	}
\label{fig_degenerate}
\end{figure}

There are two intrinsically different approaches to realize an NLI. 
The \emph{degenerate} scheme employs two parametric amplifiers (the source and the analyzer) that act as single-mode squeezers.  
Alternatively, two-mode squeezers are used in the the \emph{nondegenerate} scheme. 
Each type of NLI has its own theoretical description, which in turn has an impact on the overall phase sensitivity, as already pointed out by~\cite{Yurke86}, even though the scaling behavior is very similar.
In this article, we mostly focus on the degenerate NLI shown in Fig.~\ref{fig_degenerate} and use the subscript $\text{d}$ for all quantities derived for the degenerate case. 
For a treatment of the nondegenerate setup we refer to appendix~\ref{app_Non-degenterate_configuration}. 
For completeness, we discuss in Sec.~\ref{sec_Comparison} the results of the nondegenerate NLI and compare them to the degenerate case.

An NLI as shown in Fig.~\ref{fig_degenerate} consists of a source, which we call crystal A, that generates squeezed vacuum through parametric down-conversion. 
Its output is transmitted to an analyzing crystal, called crystal B, where it is either further squeezed or un-squeezed, depending on the phase accumulated between the crystals and by the pump.
The output of the analyzer is then detected by detector D$_\text{d}$.
For more details on the derivations of this section, we refer to appendix~\ref{app_Degenerate_configuration}.

We assume that the pair generation in each crystal A and B is described by a Bogoliubov transformation 
$
\hat{a}'= u_\text{A} \, \hat{a} + v_\text{A} \, \hat{a}^\dagger
$
and
$
\hat{b}= u_\text{B} \, \hat{a}'' + v_\text{B} \, \hat{a}^{\prime\prime\dagger}
$.
Here, $u_\text{A,B}$ and $v_\text{A,B}$ are complex parameters that describe the amplification process.
They are connected to the usual hyperbolic functions and fulfill the relation
$
1= |u_\text{A,B}|^2-|v_\text{A,B}|^2\equiv U_\text{A,B}-V_\text{A,B} 
$.
The operators $\hat{a}$ and $\hat{b}$ are photon annihilation operators and $\hat{a}^\dagger$ and $\hat{b}^\dagger$ are photon creation operators that follow the usual bosonic commutation relations, and the different primes describe the field at various instances of the interferometer.
Note that $V_{j}$ corresponds to the number of photons produced by an unseeded crystal.
However, in our setup crystal B will always be seeded by the output of crystal A.

The loss inside the NLI is modeled by a beam splitter S$_\text{d}$, which transforms the input modes $\hat{a}'$ and $\hat{\ell}$ via
$
\hat{a}'' = t_\text{d} \, \hat{a}' + r_\text{d} \, \hat{\ell}
$
and
$
\hat{\ell}' = t_\text{d}^* \, \hat{a}' - r_\text{d}^* \, \hat{\ell}
$
to the output modes, where $T_\text{d}\equiv |t_\text{d}|^2$ and $R_\text{d}\equiv |r_\text{d}|^2$ are the intensity transmittance and reflectivity with $R_\text{d} + T_\text{d} =1$.
We allow for complex $t_\text{d}$ and $r_\text{d}$ so that we can include phases that are accumulated inside the NLI.

With these transformations, the operator describing the output field of crystal B (and therefore neglecting detection loss for the moment) takes the form
\bg\begin{split}\label{e_a_B}
\hat{b}=& \, ( t_\text{d}u_\text{A} u_\text{B} +t_\text{d}^* v_\text{A}^* v_\text{B})\hat{a} +  (t_\text{d}v_\text{A} u_\text{B} +t_\text{d}^* u_\text{A}^* v_\text{B})\hat{a}^\dagger \\
& + r_\text{d} u_\text{B} \hat{\ell} + r_\text{d}^* v_\text{B} \hat{\ell}^\dagger.
\end{split}
\eg
For a vacuum input in modes $\hat{a}$ and $\hat{\ell}$ it is relatively easy to see that the photon number $N_\text{d}(\phi)\equiv\langle\hat{b}^\dagger \hat{b}\rangle$ after crystal B displays interference, that is,
\bg\label{eq_N_d}
N_\text{d}= T_\text{d} V_\text{A} + V_\text{B} +2 T_\text{d} V_\text{A} V_\text{B} - 2 T_\text{d} \sqrt{U_\text{A} V_\text{A} U_\text{B} V_\text{B}} \cos \phi.
\eg
Here we define the phase as $
\phi \equiv \operatorname{arg}\left(u_\text{A} v_\text{A} u_\text{B} v_\text{B}^*t_\text{d}^2\right)+\pi
$.
It includes the phase of the coefficients $u_j$ and $v_j$ and therefore the phase  difference of the laser field that pumps crystal A and B.
The argument of $t_\text{d}$ accounts for the phase accumulated by the photons inside the NLI.
We see that internal loss leads to a decreasing visibility and by that sensitivity, whose scaling might change for a decohering quantum state inside the interferometer~\cite{Demkowicz14}.

The variance of the photon number $\hat{b}^\dagger \hat{b}$ after crystal B may be written as (see appendix~\ref{app_Degenerate_configuration})
\bg\label{e_Delta_N}
\Delta N_\text{d}^2= 2 N_\text{d} (1+ N_\text{d}) - R_\text{d} T_\text{d} V_\text{A}.
\eg
Therefore, the photon statistics in the output of the NLI is, at least for no loss, super-thermal and, due to the fact that $N_\text{d}=N_\text{d}(\phi)$, phase dependent.
Note that $N_\text{d}$ depends on $V_\text{A}$ and $V_\text{B}$ as well as on internal loss.

The phase uncertainty of the NLI without taking detection loss into account is defined as
\bg\label{eq_sensitivity}
\Delta \phi_\text{d}^2 = \Delta N_\text{d}^2\Big/\left|\frac{\partial N_\text{d}}{\partial\phi}\right|^2
\eg
and depends on the phase $\phi$, internal loss $T_\text{d}$, as well as the gains through $V_\text{A}$ and $V_\text{B}$.
Equation~\eqref{eq_sensitivity} is the measure for the phase uncertainty that is usually employed~\cite{Yurke86,Marino12,Sahota15,Szigeti17}.
It implies that the average value of the detected photon number is used to estimate the phase~\footnote{
A phase estimator based on photon the number $N_\text{d}=\mathcal{A}-\mathcal{K}\cos\phi$, where $\mathcal{A}$ and $\mathcal{K}$ are defined in accordance with Eq.~\eqref{eq_N_d}, can be defined with the average value $n_p$ after $p$ measurements of $N_\text{d}$ as $\Phi \equiv \arccos [(\mathcal{A}-n_p)/\mathcal{K}]$. Equation~\eqref{eq_sensitivity} then scales additionally with $p^{-1}$. The estimator is, assuming the validity of the central limit theorem, asymptotically unbiased~\cite{Pezze14}
} and a more detailed motivation based on error propagation can be found in~\cite{Pezze08}.
Since only the average photon number is determined, Eq.~\eqref{eq_sensitivity} can be asymptotically linked to the Fisher information if the central limit theorem holds~\cite{Pezze14}.
We see from Eq.~\eqref{e_Delta_N} that in a balanced setup without losses, the variance is not finite and discuss this case separately in the example given below.

We emphasize that other estimators are possible and might even be a better choice than the average value of the photon number, but they might also require further information to be meaningful.
In fact, detecting the statistical properties of the signal might lead to a better estimation of the phase~\cite{Pezze08}.

To model detection loss we introduce, according to Fig.~\ref{fig_degenerate}, a second beam splitter S$_{\eta_\text{d}}$ with transmittance $\eta_\text{d}$ before the detector D$_\text{d}$.
With Eq.~\eqref{eq_sensitivity} we demonstrate in appendix~\ref{app_Degenerate_configuration} that the phase uncertainty including detection loss takes the form
\bg\label{eq_sensi_eta}
\Delta \phi_{\eta\text{d}}^2 =\Delta \phi_{\text{d}}^2 \left( 1 + \frac{1-\eta_\text{d}}{\eta_\text{d}} \frac{N_\text{d}}{\Delta N_\text{d}^2}\right).
\eg
Hence, the phase sensitivity is modified in the presence of detection loss.
In particular, it depends on the inverse Fano factor $N_\text{d}/\Delta N_\text{d}^2$, the ratio of photon number and its variance.
Therefore, the photon statistics is crucial in determining the influence of detection loss.
It is obvious from Eq.~\eqref{eq_sensi_eta} that detection loss is suppressed if the inverse Fano factor is small.
Note that a similar expression for the nondegenerate case was derived in~\cite{Marino12,Sparaciari16} for the sum of the signal of the two output ports in the nondegenerate NLI.
We discuss the limitations for this specific case in Sec.~\ref{sec_Comparison}.
An expression for the phase sensitivity in the degenerate case for equal internal and detection loss was analzed in~\cite{Sahota15}.

\paragraph*{Example: Balanced gain}

In the original work~\cite{Yurke86} the gains in the two crystals were balanced, i.e., $V_\text{A}=V_\text{B}\equiv V$ as well as $U_\text{A} = U_\text{B} \equiv U$, and no internal loss was considered, thus setting $R_\text{d}=0$.
Hence, we find from Eq.~\eqref{eq_N_d} the form $N_\text{d}= 2 UV(1-\cos\phi)$.
From Eq.~\eqref{eq_sensitivity} we obtain
\bg\label{e_sensi_equal}
\left.\Delta \phi_\text{d}^2\right|_{\phi =0}= \left.\frac{1 + 2 UV (1-\cos\phi)}{UV (1+\cos\phi)} \right|_{\phi=0}= \frac{1}{2 U V}.
\eg
Note that $U=1+V$, and that $V$ corresponds to the number of photons that are produced by crystal A and are annihilated by crystal B.
Because $V$ photons are inside the NLI and interact with a possible object, it is said that the NLI has a Heisenberg scaling of the phase sensitivity.
The choice of $\phi=0$ corresponds to the phase where the phase uncertainty $\Delta \phi_\text{d}^2$ is minimal~\cite{Yurke86}.
Therefore, the NLI would be ideally operated at this point.
We show in appendix~\ref{app_Quantum_Fisher_information} that the quantum Fisher information is $2UV$ and therefore, Eq.~\eqref{e_sensi_equal} saturates the quantum Cram\'er-Rao bound.

However, with equal gains and for this phase all photons created by the source are annihilated by the analyzer and we have $N_\text{d}|_{\phi=0}=0$, i.\,e., we expect to measure no photons in the output of the NLI.
This fact is particularly unfavorable because it means that vacuum fluctuations are of the same order of magnitude.
Since in a realistic experiment these fluctuations are introduced by non-perfect detectors---in our treatment modeled by S$_{\eta_\text{d}}$---they significantly reduce the phase sensitivity.

\begin{figure}[htb] \centering
\includegraphics[scale=.99]{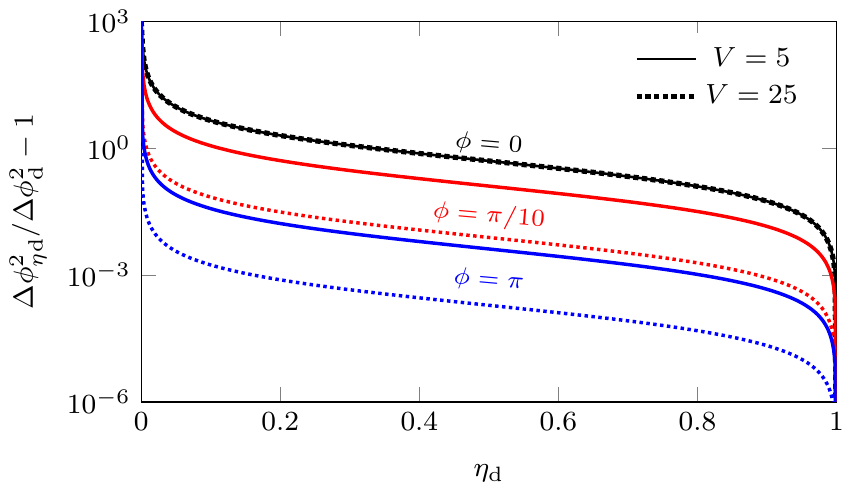}
    \caption{
   Relative deviation of the phase uncertainty under the influence of detection loss $\eta_\text{d}$ from the lossless case for equal gain. We show Eq.~\eqref{e_det_loss_equal} for three different phases $0$, $\pi/10$ and $\pi$ (black, red and blue) for two different gain parameters $V=5$ and $V=25$ (solid and dashed). At destructive interference, the influence of detection loss is the highest and is always the same, independent of the gain. 
    }
\label{fig_det_loss}
\end{figure}
The effect becomes obvious when we note that the inverse Fano factor $N_\text{d}/\Delta N^2_\text{d} = 1/(2+2N_\text{d})$ and, following Eq.~\eqref{eq_sensi_eta}, we arrive at
\bg\label{e_det_loss_equal}
\frac{\Delta \phi_{\eta\text{d}}^2}{\Delta \phi_{\text{d}}^2}-1 =\frac{1-\eta_\text{d}}{2\eta_\text{d}}\frac{1}{1+2UV(1-\cos\phi)} .
\eg
To provide a quantitative analysis of the relative deviation of the phase uncertainty from the uncertainty without detection loss, we plot Eq.~\eqref{e_det_loss_equal} in Fig.~\ref{fig_det_loss} as a function of detection loss $\eta_\text{d}$.
The deviation depends on the phase, the loss, and the gain.
For constructive interference ($\phi=\pi$), we see that the deviation is the smallest and even further reduced with increasing gain.
At destructive interference ($\phi=0$), we have a deviation that has a similar functional behavior, but is orders of magnitude larger than for constructive interference.
Moreover, increasing the gain does not decrease the deviation (the black solid and dashed lines overlap).
In fact, it can be easily seen that Eq.~\eqref{e_det_loss_equal} reduces to $\Delta \phi^2_{\eta \text{d}} = \Delta \phi_\text{d}^2 (1+\eta_\text{d})/(2\eta_\text{d})$ for $\phi=0$, in total agreement with the expression for the phase uncertainty of the sum of the two output ports in the nondegenerate NLI discussed in~\cite{Marino12,Sparaciari16}.
If we were not to operate this NLI at or close to destructive interference, we could get a significant number of photons exiting the device and therefore suppress the influence of the detection loss.
However, the minimal phase uncertainty occurs exactly at vanishing $\phi$ and with it at vanishing $N_\text{d}$, and only in this case do we obtain the unique Heisenberg scaling of the uncertainty.

As this example demonstrates, the effect of detection loss in an unseeded NLI is governed by the intensity $N_\text{d}(\phi)$ in the output of the interferometer.
Since we are mainly interested in the phase where its uncertainty is minimal, we do not have the flexibility to operate the interferometer at a different phase, e.g., at constructive interference where $N_\text{d}(\phi)$ is maximal.
However, there is a different option for increasing $N_\text{d}(\phi)$, namely, using different gain values for the two crystals, that is, unbalancing the gains.
If, for example, the source is stronger than the analyzer, all the photons created in crystal A can never be annihilated in crystal B, even if the interferometer is set to destructive interference.
In the opposite case in which the analyzer is weaker than the source, crystal B not only annihilates all photons emerging from crystal A, but overcompensates and creates additional photons.
Therefore, the larger the gain difference, the higher the intensity in the output of the interferometer and the smaller the impact of detection loss.
Hence, we expect a suppression of detection loss for a gain-unbalanced setup~\cite{Sparaciari16,Manceau17,Manceau17PRL}.

\section{Phase sensitivity for unbalanced gain}
\label{sec_Phase_sensitivity_for_unbalanced_gain}

In the section above we established that unbalancing the gain can be beneficial if there is significant detection loss in the NLI.
However, even though the impact of detection loss is reduced, the effect of unbalanced gain on the phase uncertainty itself has not been studied yet.
In this section, we derive the minimal phase uncertainty for vanishing internal loss $R_\text{d}=0$.
For that, we minimize Eq.~\eqref{eq_sensitivity} for arbitrary gains and find that the minimum  uncertainty occurs at the phase
\bg\label{e_phi_min}
\phi_\text{min}=\pm \arctan\sqrt{\frac{V_\text{max} U_\text{max}- V_\text{min} U_\text{min}}{(U_\text{max}+V_\text{max})^2 V_\text{min} U_\text{min} }},
\eg
which leads with Eq.~\eqref{eq_N_d} to the number of photons
\bg
N_\text{d} (\phi_\text{min})= \frac{V_\text{max}-V_\text{min}}{U_\text{min}+V_\text{min}}.
\eg
Here, we defined $V_\text{min} = \operatorname{min}[V_\text{A},V_\text{B}]$ and $V_\text{max}=\operatorname{max}[V_\text{A},V_\text{B}]$ as the smaller and the larger parameter, respectively.
The parameters $U_\text{min}$ and $U_\text{max}$ are defined in an analogous way.
Moreover, we see that $ N_\text{d} (\phi_\text{min})\gg 1$ if $V_\text{min}\ll V_\text{max}$.
In this case, detection loss has practically no impact on the phase uncertainty, which one can directly see from Eq.~\eqref{eq_sensi_eta}.
Note further that for $V_\text{A}\neq V_\text{B}$ the optimal sensitivity is  achieved if the NLI is not operated at destructive interference, $\phi_\text{min}\neq0$.

With the phase from Eq.~\eqref{e_phi_min} we find that the minimal phase uncertainty takes the form
\bg\label{e_Delta_phi_d_min}
\Delta \phi^2_\text{d}(\phi_\text{min}) = \frac{1}{2 U_\text{min}V_\text{min} }.
\eg
Hence, the phase sensitivity is limited by the crystal with \emph{smaller} gain, independent on whether it is crystal A or B.
In the numerical analysis of~\cite{Sparaciari16} it was implicitly seen that the smaller gain limits the sensitivity, but this fact was not commented upon further.

\paragraph*{Quantum limitations}

The attention that NLIs have attracted is due to the scaling behavior of their phase uncertainty, which is often referred to as the `Heisenberg scaling.'
Indeed, we saw from Eq.~\eqref{e_sensi_equal} in a gain-balanced NLI with $V_\text{A}=V_\text{B}=V\gg1$ that $\Delta \phi_\text{d} (0)\cong 1/(\sqrt{2}V)$.
Since $V$ corresponds to the number of photons produced by the source, the connection to the Heisenberg scaling is evident.
However, a disadvantage of a gain-balanced NLI is that it is very susceptible to detection loss.

In an unbalanced setup at high gain, we find from Eq.~\eqref{e_Delta_phi_d_min} that $\Delta \phi_\text{d} (\phi_\text{min})\cong 1/(\sqrt{2}V_\text{min})$, where $V_\text{min}$ is the smaller gain parameter.
If the source is weaker than the analyzer ($V_\text{A}< V_\text{B}$), the sensitivity is limited by $V_\text{A}$ and therefore by the number of photons that interact with the object.
A comparison to the shot-noise level of this photon number seems obvious and the phase sensitivity indeed displays a Heisenberg scaling.
Such phase measurements below the shot-noise level determined by $V_\text{A}$ have recently been performed using direct detection~\cite{Manceau17PRL}.

If the analzyer is weaker than the source ($V_\text{B}<V_\text{A}$), the sensitivity is limited by $V_\text{B}$, which is completely independent of how many photons interacted with the object or were inside the NLI, described by $V_\text{A}$.
However, it is $V_\text{A}$ that is conventionally used~\cite{Sparaciari16} to determine a shot-noise or Heisenberg scaling behavior.
In this case the comparison would be somewhat artificial, because the sensitivity is not limited by this number.

Of course, if the sample in the interferometer is very sensitive and gets easily destroyed by high intensities or if radiation pressure on mirrors degrades the sensitivity, one would always operate the NLI with the smaller number of photons inside and naturally choose $V_\text{A}<V_\text{B}$ so that the analyzer is stronger.
But if there is no limitation on how many photons might interact with an object inside the NLI, there is no preference as to which of the the two gains should be the lower one because the resulting sensitivity is exactly the same.
The interferometer is completely symmetric and the only limiting factor is the crystal with smaller gain, independent of which crystal it is.
Hence, a comparison to a `Heisenberg limit' is not straightforward and has to be justified in each case, let alone the fact that the pump is assumed to be undepleted.
In fact, other variations of an NLI give a phase sensitivity that scales with the shot-noise level of pump photons~\cite{Szigeti17}.

In conclusion, the second crystal has to be considered an essential part of the interferometer.
Of course it is valid to employ truncated schemes~\cite{Anderson17} if the output of an NLI is detected by homodyne detection~\cite{Hudelist14}, but the original proposal~\cite{Yurke86} only involves a much simpler direct detection scheme~\cite{Linnemann16,Manceau17PRL}, in which the analyzing crystal is vital.
On the same note, the quantum Fisher information and the quantum Cram\'er-Rao bound do not specify a particular detection scheme and are usually calculated for the state inside the interferometer~\cite{Sparaciari15,Sparaciari16,Anderson17b,Anderson17}.
Because the analyzer is seen as a part of the detection, the bound is independent of the gain of crystal B and since one optimizes over all possible detection schemes, it is implicitly assumed that this gain can be arbitrarily high.
On the other hand, if one sees crystal B as an integral component of the interferometer, one cannot just optimize over all possible parameters but is restricted by the experimental limitations.

\section{Breaking the symmetry through internal loss}
\label{sec_Breaking_the_symmetry}
In the previous section we pointed out that the  phase sensitivity is limited by the lower gain, and therefore is completely symmetric with respect to the source and analyzer.
If we introduce internal loss, this symmetry is broken.

It is straightforward to see that including internal loss, i.e., a beam splitter S$_\text{d}$ with $T_\text{d}< 1$ in-between the two crystals as shown in Fig.~\ref{fig_degenerate}, affects solely the photons created in crystal A and only indirectly through a modified input the action of crystal B.
This fact stands out most clearly by observing that in both Eqs.~\eqref{eq_N_d} and \eqref{e_Delta_N} the quantity $V_\text{A}$ always appears together with the transmittance $T_\text{d}$.
For simplicity we introduce the notation
$
V_t \equiv T_\text{d} V_\text{A},
$
which describes the number of photons that are transmitted from crystal A to crystal B.
In agreement with the previous notation we also use $U_t \equiv 1 + V_t$.
In the following we first give an intuitive explanation of the influence of internal loss on the sensitivity, before turning to the exact results.

If internal loss is small, we can neglect the term $- R_\text{d} T_\text{d}V_\text{A}$ in Eq.~\eqref{e_Delta_N} for the variance.
Moreover, if operated at high gain, $V_t \gg 1  $, the photon number as described in Eq.~\eqref{eq_N_d} takes the form
\bg
N_\text{d}(\phi)\cong V_t + V_\text{B} + 2 V_t V_\text{B} - 2 \sqrt{U_t U_\text{B} V_t V_\text{B}} \cos \phi
\eg
and therefore we arrive at the same result as in Eq.~\eqref{e_Delta_phi_d_min} for the minimal phase uncertainty, with only $V_\text{A}$ replaced by $V_t$.
In fact, an expansion of the exact treatment given below in orders of $R_\text{d}$ gives rise, up to lowest order, to the phase uncertainty $\Delta\phi^2 \cong 1/ (2 U_\text{min} V_\text{min})$, where now $V_\text{min} = \operatorname{min}[V_t,V_\text{B}]$ and $V_\text{max}=\operatorname{max}[V_t,V_\text{B}]$.
Because $V_t$ depends on $T_\text{d}$, the phase uncertainty is independent of the transmittance only if $V_\text{B} < V_t$.
Thus, if internal loss dominates it is better to have a stronger source than analyzer, the opposite case of what was used in~\cite{Sparaciari16,Manceau17,Manceau17PRL}.

\begin{figure}[htb]
\includegraphics[scale=.98]{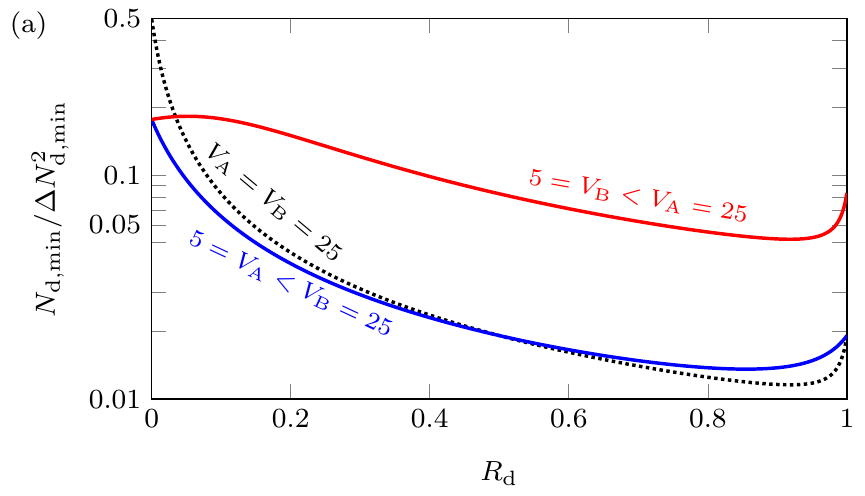} \includegraphics[scale=.98]{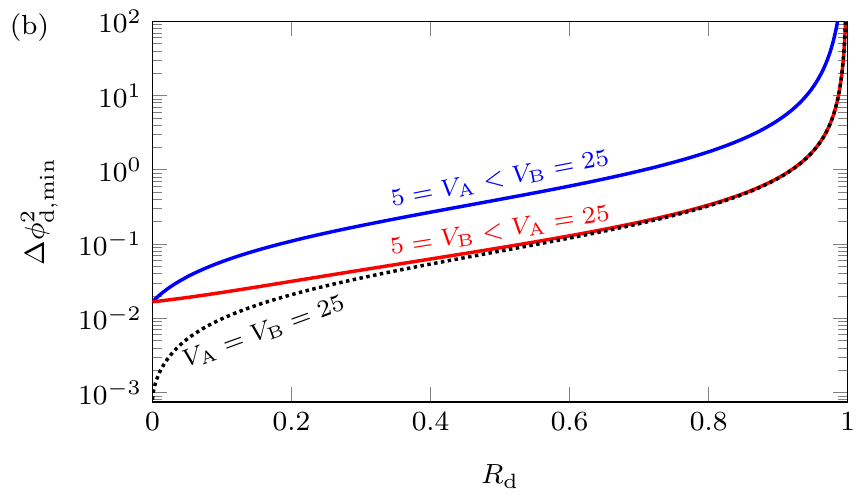}
    \caption{
    Effect of internal loss $R_\text{d}$ on  $N_\text{d,min}/\Delta N^2_\text{d,min}$ in part (a) and on $\Delta\phi^2_\text{d,min}$ in part (b).
    In part (a) we demonstrate that the inverse Fano factor $N_\text{d}/\Delta N_\text{d}^2$ decreases rapidly for the gain-balanced situation (dotted line), because internal loss deteriorates the number of transmitted photons $V_t$.
	Therefore, the gain of crystal A is effectively lower and detection loss is suppressed.
    The same is true if the analyzer is stronger than the source (blue solid line) and the advantage of unbalancing the gain decreases.
    If the source is stronger (red solid line), the inverse Fano factor increases slightly at first.
    Moreover, it is always larger than for the opposite case.
    In part (b) we show that the phase uncertainty (without detection loss) is the smallest for a gain-balanced setup (dotted line).
	For large loss $R_\text{d}$, the case of a stronger source (red line) is very close to the gain-balanced result.
	Moreover, for an unbalanced setup it is always beneficial to work with a stronger source than with a stronger analyzer (blue line).
    }
\label{fig_internal_loss}
\end{figure}

However, for a more accurate description of the NLI we derive from \eqref{eq_sensitivity} an expression for the phase where its uncertainty is minimized.
We can perform this calculation analytically, but the expressions are rather cumbersome and we therefore refrain from presenting them.
When we use this phase in Eq.~\eqref{eq_N_d} to calculate the photon number, we find
\begin{widetext}
\bg\label{e_N_loss}
N_\text{d,min}=
\frac{2 (V_\text{B}-V_t)^2 +  \mathcal{L}+ \sqrt{4 (V_\text{B}-V_t)^2 (U_\text{B} +V_t)^2 +4  U_\text{B}V_\text{B} \mathcal{L} +R_\text{d} V_t [2 U_\text{B} V_\text{B} -U_t V_t]+ \mathcal{L}^2 }}{2(U_\text{B}+V_\text{B})(U_t+V_t)},
\eg
\end{widetext}
where we defined a loss-dependent term $\mathcal{L} \equiv R_\text{d} V_t (1 + 8 U_\text{B} V_\text{B}) $.
With this analytic expression we are able to determine the variance $\Delta N^2_\text{d,min} = 2 N_\text{d,min} (1+N_\text{d,min})- R_\text{d} V_t$ from Eq.~\eqref{e_Delta_N}. 
The inverse Fano factor $N_\text{d,min}/\Delta N^2_\text{d,min}$ that suppresses the effect of detection loss according to Eq.~\eqref{eq_sensi_eta} can be calculated for different parameters.
We plot this factor in Fig.~\ref{fig_internal_loss}\,(a) on a logarithmic scale.
For $R_\text{d}=0$ unbalancing the gains decreases the influence of detection loss significantly: the dotted line describing the balanced configuration is much higher than the red and blue solid lines with a stronger analyzer and source, respectively.
However, the number of photons transmitted to crystal B, $V_t = (1-R_\text{d}) V_\text{A}$, decreases for $R_\text{d}>0$ and for an initially balanced situation we arrive effectively at a gain-unbalanced setup with $V_t < V_\text{B}$.
Hence, the dotted line decreases rapidly until it is very close to the case of a stronger analyzer (blue line).

For the same reason, a stronger source at first slightly increases the inverse Fano factor because the unbalancing is effectively lowered, making detection loss more significant again.
On the other hand, for a stronger analyzer we see a decrease of the inverse Fano factor, because due to internal loss the gain-unbalancing effectively increases and we have $V_t< V_\text{A} < V_\text{B}$.
Hence, detection loss is further suppressed and in this sense the setup improves.
The plot also demonstrates that it is better to have a stronger analyzer to suppress the effect of detection loss in the presence of internal loss, which is the configuration that was investigated in~\cite{Sparaciari16,Manceau17,Manceau17PRL}.
Note further that only for $R_\text{d}\cong 0 $ we see that there is a significant advantage of gain-unbalancing and in this case it does not matter much which one of the crystals has higher gain.

After considering the inverse Fano factor $N_\text{d}/\Delta N_\text{d}^2$ that suppresses detection loss, we now turn to the minimal phase uncertainty itself.
We find the analytical expression
\bg\label{e_delta_phi_loss}
\begin{split}
\Delta \phi_\text{d,min}^2 = &[N_\text{d,min}(U_\text{B} + V_\text{B}) (U_t+V_t)+ U_\text{B} V_t + \\
&U_t V_\text{B} - R_\text{d} V_t][4 U_\text{B} V_\text{B} (U_t-R_\text{d}) V_t]^{-1}
\end{split}
\eg
for the phase uncertainty without detection loss and plot it in Fig.~\ref{fig_internal_loss}\,(b).
Here, the effect of internal loss is exactly opposite to the one on detection loss:
for an unbalanced situation, the phase sensitivity is always better if the source is stronger (red solid line), compared to the case where the analyzer is stronger (blue solid line).
This can be intuitively understood considering that internal loss can affect the output of crystal A directly and the action of crystal B only indirectly by modifying its input.
Hence, it is beneficial to have a stronger source so that the reduced number of photons in the interferometer is still large enough to not limit the sensitivity.

Not surprisingly, for equal gain (using the higher of the two gains for both crystals) always outperforms the unbalanced setup.
However, for sufficiently large loss, the balanced situation is very close to the case of a stronger source.

Since in a gain-unbalanced setup with internal loss a stronger analyzer suppresses the effect of detection loss, whereas a stronger source in addition reduces the influence of internal loss, the latter one seems at first sight advantageous.
However, a stronger source suppresses detection loss not as well as a stronger analyzer and is only beneficial for small internal loss when compared to the balanced setup.
Therefore, the decision which crystal to pump stronger has to be made based on the order of magnitude of the internal and detection losses.
In each individual case, it might be beneficial to have stronger source or a stronger analyzer.

\section{Comparison of degenerate and nondegenerate configuration}

\label{sec_Comparison}

\begin{figure}[htb] \centering
\includegraphics[scale=1]{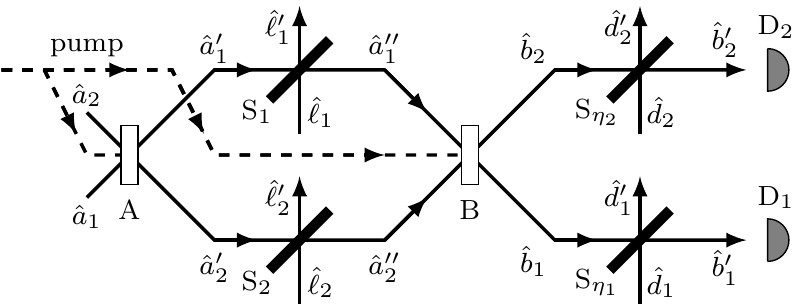}
    \caption{
    Schematic of a nondegenerate nonlinear interferometer consisting of two coherently pumped nonlinear crystals A and B.
    Internal loss is modeled by a beam splitter S$_\text{1,2}$ in each arm, detection loss by a beam splitter S$_{\eta_{1,2}}$ in front of each detector D$_{1,2}$.
	}
\label{fig_nondegenerate}
\end{figure}

So far we have only considered a degenerate setup.
However, we generalize our results to the case of a nondegenerate NLI (see appendix~\ref{app_Non-degenterate_configuration}) and compare them in this section to the degenerate ones obtained above.
The setup is shown in Fig.~\ref{fig_nondegenerate}, where we now have different input and output modes $1$ and $2$.
Even though the expressions for the photon number and variance are different from the degenerate configuration, they are similar enough so that our previous discussion can also be applied---with some limitations---to the nondegenerate case. 

The photon number detected by detector D$_j$ in a nondegenerate as well as a degenerate configuration takes the form
\bg
N_{\eta j}= \eta_j N_j =\eta_j  \left( \mathcal{A}_j - \mathcal{K}_j \cos \phi \right),
\eg
where $\mathcal{A}_j$ is the amplitude and $\mathcal{K}_j$ the contrast of the signal and the index $j=\text{d},1,2$.
Here, $\eta_j$ is the efficiency of the detector $j$ modeled by a beam splitter S$_{\eta j}$ and $T_j$ the transmittance of the beam splitter S$_j$ between the two crystals.
$R_j$ is the corresponding reflectivity.
The explicit form of $\mathcal{A}_j$ and $\mathcal{K}_j$ is summarized in table~\ref{t_comparison} for all cases.
The internal loss in each arm of the NLI may be different and has an effect on the signal and variance~\cite{Xin16}.
Note that $\phi$ is defined slightly different from the degenerate case to account for the different phases in the two branches of the NLI.

\begin{table*}
\caption{\label{t_comparison}
Comparison of degenerate and nondegenerate NLI.
The index $\text{d}$ denotes the degenerate case, $1$ and $2$ the two output ports of the nondegenerate case, and $+$ their sum.
The signal without detection loss has an amplitude $\mathcal{A}_j$ and a contrast $\mathcal{K}_j$.
The variance of the detected signal is $\Delta N_{\eta j}^2$.
In the second part of the table we show the variance $\Delta N_{j}^2$ without detection loss and later the same quantity for vanishing internal loss.
The factor in the second column shows the influence of detection loss in a balanced situation.
The optimal phase uncertainty $\Delta \phi_j^2(\phi_\text{min})$ was calculated for the lossless case.
}
\footnotesize\rm
\begin{tabular*}{\textwidth}{@{}l*{15}{@{\extracolsep{0pt plus1pt}}l}}
\hline\hline
\addlinespace[.4ex]
$j$& $\mathcal{A}_j$ & $\mathcal{K}_j$ & $\Delta N_{\eta j}^2$ \\
\hline
\addlinespace[.4ex]
$\text{d}$ & $T_\text{d} V_\text{A}+ V_\text{B} + 2 T_\text{d} V_\text{A} V_\text{B} $& $ 2T_\text{d}\sqrt{U_\text{A} U_\text{B} V_\text{A} V_\text{B}}$&$\eta_\text{d} N_\text{d} (1+ \eta_\text{d} + 2 \eta_\text{d} N_\text{d})- \eta_\text{d}^2 R_\text{d} T_\text{d} V_\text{A}$ \\
\arrayrulecolor{gray}\hline
\addlinespace[.4ex]
$1$&$T_1 V_\text{A} + V_\text{B} +(T_1+T_2)V_\text{A} V_\text{B} $ & $ 2\sqrt{T_1 T_2 U_\text{A} U_\text{B} V_\text{A} V_\text{B}}   $ & $\eta_1 N_1 (1+ \eta_1 N_1)$ \\
$2$&$T_2 V_\text{A} + V_\text{B} +(T_1+T_2)V_\text{A} V_\text{B} $ & $ 2\sqrt{T_1 T_2 U_\text{A} U_\text{B} V_\text{A} V_\text{B}} $&$\eta_2 N_1 (1+ \eta_2 N_2)$ \\
$+$&$\mathcal{A}_1+\mathcal{A}_2  $& $ 4\sqrt{T_1 T_2 U_\text{A} U_\text{B} V_\text{A} V_\text{B}} $ &  $(\eta_1 N_1 + \eta_2 N_2) (1+ \eta_1 N_1+ \eta_1 N_1) $\\
&&& $ + \eta_1 \eta_2 (N_1+ N_2) - \eta_1 \eta_2 (T_1+T_2-2T_1 T_2)V_\text{A} $ \\
\arrayrulecolor{black}
\hline
\end{tabular*}
\begin{tabular*}{\textwidth}{@{}l*{15}{@{\extracolsep{0pt plus1pt}}l}}
\hline
\addlinespace[.4ex]
$j$& $\Delta N_{ j}^2$& $\left.1 + \frac{1-\eta_j}{\eta_j} \frac{N_j}{\Delta N_j}\right|_{\phi=0}$& $\Delta N_j^2$ with $T_j=1$ &  $\Delta \phi_j^2(\phi_\text{min})$  \\
\hline
\addlinespace[.4ex]
$\text{d}$& $ 2 N_\text{d} (1+ N_\text{d}) - R_\text{d} T_\text{d} V_\text{A}  $& $(1+\eta_\text{d})/(2\eta_\text{d})$ & $2 N_\text{d} (1+N_\text{d})$& $1/(2 U_\text{min} V_\text{min})$  \\
\arrayrulecolor{gray}\hline
\addlinespace[.4ex]
$1$ & $ N_1 (1+ N_1)$& $1/\eta_\text{1}$  & $ N_\text{d} (1+N_\text{d})$& $1/(4 U_\text{min} V_\text{min})$ \\
$2$ & $ N_2 (1+ N_2)$&  $1/\eta_\text{2}$  & $ N_\text{d} (1+N_\text{d})$& $1/(4 U_\text{min} V_\text{min})$ \\
$\text{+}$ & $ N_+ (2+ N_+) +  [2 T_1 T_2 - (T_1+T_2)]V_\text{A}$& $(1+\eta_+)/(2\eta_+)$  & $4 N_\text{d} (1+N_\text{d})$& $1/(4 U_\text{min} V_\text{min})$  \\
\arrayrulecolor{black}
\hline\hline
\end{tabular*}
\end{table*}

In the nondegenerate setup one can, in addition to considering both exit ports separately, analyze the sum of the two signals~\cite{Yurke86}.
Without detection loss, we therefore define the sum of the two signals $N_+ \equiv N_1+N_2$.
Ultimately, we are interested in the phase uncertainty $\Delta \phi_j^2$ which can be analogously defined to Eq.~\eqref{eq_sensitivity} where we replace the index $\text{d}$ by $j=1,2,+$.
To obtain the phase uncertainty, we first need the variance of the photon number.

We display in the table the variances $\Delta N_{\eta j}^2$ and $\Delta N_j^2$ with and without detection loss, respectively, and note that only for $j=\text{d},1,2$ the relation
\bg\label{e_Delta_N_j_eta}
\Delta N_{\eta j}^2 = \eta_j^2 \Delta N_j^2 + \eta_j (1-\eta_j) N_j
\eg
holds, but not for the sum of the two signals.
Therefore, the suppression of the detection loss for phase sensitivity of the signal-sum is not as straightforward.
However, if $\eta_1 = \eta_2 \equiv \eta_+$, we also find Eq.~\eqref{e_Delta_N_j_eta} for $j=+$ as predicted by~\cite{Marino12}.
In analogy to appendix~\ref{app_Degenerate_configuration} we arrive at 
\bg
\Delta \phi_{\eta j }^2 = \Delta \phi^2_j \left( 1 + \frac{1-\eta_j}{\eta_j} \frac{N_j}{\Delta N_j^2}\right)
\eg
and thus that the inverse Fano factor $N_j/\Delta N_j^2$ determines the suppression of detection loss with $j= \text{d},1,2,+$.
In fact, for a gain-unbalanced scheme and vanishing internal loss we find for the expression in parentheses the factor shown in table~\ref{t_comparison} in agreement with~\cite{Marino12}.
Note that if only one single exit port is detected in the nondegenerate setup, we find a significantly different dependence on detection loss.
We also see that gain-unbalancing suppresses detection loss in the nondegenerate setup, even though for the sum of the two signals and $\eta_1\neq \eta_2$ the treatment is more subtle.

If the internal loss in each arm is equal, i.e., $T_1=T_2=T_\text{d}$, we find $N_1=N_2=N_\text{d}$ and $N_+ = 2 N_\text{d}$.
Moreover, if we assume the lossless case with $T_j=1$, we can express all variances simply by the number of photons in one exit port $N_\text{d}$ (see the third column in the lower part of table~\ref{t_comparison}).
Since all of these results differ only by a factor from the degenerate result, we can use the phase from Eq.~\eqref{e_phi_min} to obtain the minimal uncertainty.
These results are shown in the table as well.
We note that the phase uncertainty is smaller by a factor of two in the nondegenerate case, regardless of using only one detector or both.
In particular, we obtain again that the sensitivity is limited by $V_\text{min}=\operatorname{min}[V_\text{A},V_\text{B}]$.
Hence, we see that the discussion from above similarly applies to the different cases of a nondegenerate setup.
This is also true if we include internal loss in analogy to Sec.~\ref{sec_Breaking_the_symmetry}, even though there are more parameters since there might be different loss in each arm of the NLI.
We therefore refrain from presenting a lengthy discussion of all different cases.

Finally, to compare the sensitivity to the shot-noise level in a nondegenerate setup, one has to remember that $V_\text{A}$ is the number of photons per mode produced by crystal A.
Therefore, the number of photons inside the NLI is $n_+ = 2 V_\text{A}$, whereas in the degenerate case it was $n_\text{d}= V_\text{A}$.
Hence, the shot-noise level could be defined as $1/\sqrt{2V_\text{A}}$ in contrast to the degenerate NLI, where it is $1/\sqrt{V_\text{A}}$.
In case of a stronger analyzer, we have $\Delta \phi_\text{d,min} \cong 1/(\sqrt{2}V_\text{A})=1/(\sqrt{2}n_\text{d})$ and $\Delta \phi_{+\text{,min}} \cong 1/(2V_\text{A})=1/n_+$.
In absolute values, $\Delta \phi_{+\text{,min}}<\Delta \phi_\text{d,min}$.
However, if we assume the same number of photons inside the interferometer, that is, $n_\text{d}= n_+$, we find $\Delta \phi_\text{d,min}=\Delta \phi_{+\text{,min}}/\sqrt{2}$ in contrast to our previous statement.

\section{Conclusions}
\label{sec_Conclusions}

We have demonstrated that the sensitivity of a degenerate NLI is limited by the crystal with the smaller gain, whether it is the source or analyzer crystal.
Hence, the second crystal has to be considered an essential part of the interferometer and its gain is equally important as the one of the source in a setup without internal loss.
Moreover, the sensitivity might not scale at all with the number of photons produced by the source.
We emphasize that a comparison to the shot-noise or Heisenberg limit is only suggestive if the gain of the source is the limiting factor.
If the analyzer is limiting the sensitivity, a comparison to the shot-noise level and a discussion of a `Heisenberg scaling' seems rather artificial.
Together with the discussion of~\cite{Szigeti17}, we therefore hope to raise awareness for the subtleties of claiming a Heisenberg scaling.

In order to suppress the effect of detection loss, it might be beneficial to unbalance the gains of the two crystals on purpose.
Indeed, we showed that detection loss is suppressed by the inverse Fano factor of the photon statistics.
For a gain-balanced NLI, the optimal phase occurs for a vacuum output state and the sensitivity is susceptible to detection loss.
In contrast, unbalancing the gains leads to a significant photon number in the output that suppresses it.

Whereas for this suppression it is irrelevant whether the source or the analyzer is stronger---the NLI is symmetric in this sense---it changes dramatically when internal loss is considered.
Internal loss effectively changes the gain of the source and therefore may increase or decrease the suppression of detection loss.
In addition, this broken symmetry between the two nonlinear crystals has the consequence that a higher gain in the source reduces the effect of internal loss on the phase sensitivity.

To suppress negative effects of internal loss, a stronger source should be used; to additionally suppress detection loss, a stronger analyzer seems beneficial.
In fact, a stronger source with internal loss suppresses detection loss, but not as well as a stronger analyzer.
Hence, the decision on whether to use a higher gain for the source or for the analyzer has to be based on the magnitude of internal and detection losses for each individual case.
We emphasize that these results are valid for a degenerate NLI, but most of them carry over to the nondegenerate case, for which we provide analytical expressions as well.

\acknowledgments{
This research was performed as part of a collaboration within the Max Planck-uOttawa Centre for Extreme and Quantum Photonics, whose support we gratefully acknowledge.
We thank Maria V. Chekhova for valuable comments and discussions.
E.G., S.L., R.F., and R.W.B. are thankful for the financial support by the Canada Excellence Research Chairs program (CERC) and the Natural Sciences and Engineering Research Council of Canada (NSERC).
R.F. acknowledges the financial support of the Banting postdoctoral fellowship of the NSERC and S.L. the financial support from Le Fonds de Recherche du Qu\'{e}bec Nature et Technologies (FRQNT).
E.G. is grateful to the Friedrich-Alexander-Universit\"at Erlangen-N\"urnberg for an Eugen Lommel stipend.
}

\appendix

\section{Degenerate configuration}
\label{app_Degenerate_configuration}

In this appendix, we use the notation according to Fig.~\ref{fig_degenerate}, where the input mode of the degenerate NLI is denoted by the operator $\hat{a}$.
It describes, as its subsequent counterparts, a photonic annihilation operator and fulfills the bosonic commutation relation $[\hat{a},\hat{a}^\dagger]=1$.
The input enters at first crystal A, and its output $\hat{a}'$ is described by the Bogoliubov transformation
\bg \label{e_a_prime}
\hat{a}'=u_\text{A} \hat{a}+ v_\text{A} \hat{a}^{\dagger}.
\eg
Here, $u_\text{A}$ and $v_\text{A}$ are complex parameters.
They describe the amplification process and fulfill the relation $1= |u_\text{A}|^2-|v_\text{A}|^2=U_\text{A}-V_\text{A} $.
Due to this identity we can identify $U_\text{A}$ and $V_\text{A}$ with respective hyperbolic functions, that is $U_\text{A} = \cosh^2 r_\text{A}$ and $ V_\text{A} = \sinh^2 r_\text{A}$, where we introduced the so-called squeezing parameter or gain $r_\text{A}$ of crystal A.

The internal loss of the NLI is modeled by a beam splitter S$_\text{d}$, which is described by the transformation
\bg
\hat{a}'' = t_\text{d} \hat{a}' + r_\text{d} \hat{\ell} \quad\text{ and }\quad \hat{\ell}' = t_\text{d}^* \hat{\ell} - r_\text{d}^* \hat{a}',
\eg
where $\hat{\ell}$ is the operator associated with the noise input of the beam splitter according to Fig.~\ref{fig_degenerate} and causes vacuum noise.
We displayed also the transformation for the output $\hat{\ell}'$ to show that $t_\text{d}$ and $r_\text{d}$ may be chosen complex and there is in addition a phase shift.
The asterix ($*$) denotes the complex conjugate.
Note that $t_\text{d}$ and $r_\text{d}$ describe the field transmittance and reflectivity, respectively.
They fulfill $1= |r_\text{d}|^2 + |t_\text{d}|^2\equiv R_\text{d}+T_\text{d}$.
Choosing complex $t_\text{d}$ and $r_\text{d}$ makes it possible to absorb phases of input modes into their definition.
We therefore do not have to treat phases accumulated inside the NLI separately.
With the relation Eq.~\eqref{e_a_prime} for crystal A we find
\bg\label{e_a_prime_prime}
\hat{a}'' = t_\text{d} u_\text{A} \hat{a}+ t_\text{d} v_\text{A} \hat{a}^{\dagger} + r_\text{d} \hat{\ell}.
\eg
The action of crystal B is again described by a Bogoliubov transformation
\bg
\hat{b}=u_\text{B} \hat{a}'' + v_\text{B} \hat{a}^{\prime\prime \dagger}
\eg
with the same assumptions and notations for the coefficients $u_\text{B}$ and $v_\text{B}$ as for crystal A.
With the help of Eq.~\eqref{e_a_prime_prime} we find
\bg\label{e_app_a_prime_prime_prime}
\begin{split}
\hat{b}=& \,  (t_\text{d}u_\text{A} u_\text{B} + t_\text{d}^*v_\text{A}^* v_\text{B})\hat{a} +  (t_\text{d}v_\text{A} u_\text{B} +t_\text{d}^* u_\text{A}^* v_\text{B})\hat{a}^\dagger \\
&+ r_\text{d} u_\text{B} \hat{\ell} + r_\text{d}^* v_\text{B} \hat{\ell}^\dagger.
\end{split}
\eg
Detection loss is modeled by the transformation
\bg\label{e_app_det_loss}
\hat{b}'= \sqrt{\eta_\text{d}} \hat{b} + \sqrt{1-\eta_\text{d}} \hat{d},
\eg
which corresponds to a beam splitter S$_{\eta_\text{d}}$ in front of the detector.
Here, $\eta_\text{d}$ is the detection efficiency.

Let us assume that there is vacuum input in mode $\hat{d}$.
We then find
\bg\label{e_app_det_bs}
\hat{b}^{\prime\dagger}\hat{b}'\ket{0_d}= \eta_\text{d}\hat{b}^\dagger\hat{b} \ket{0_d} + \sqrt{\eta_\text{d}(1-\eta_\text{d})}\hat{b} \ket{1_d}
\eg
and obtain, projecting with $\bra{0_d}$ onto Eq.~\eqref{e_app_det_bs},
\begin{subequations}\label{e_app_n-eta}
\bg
\bra{0_d}\hat{b}^{\prime\dagger}\hat{b}^\prime \ket{0_d} = \eta_\text{d}\hat{b}^\dagger\hat{b}
\eg
and
\bg
\bra{0_d}\left(\hat{b}^{\prime\dagger}\hat{b}^{\prime} \right)^2\ket{0_d} = \eta_\text{d}^2 \left(\hat{b}^\dagger\hat{b}\right)^2 + \eta_\text{d}(1-\eta_\text{d})\hat{b}^\dagger\hat{b}
\eg
\end{subequations}
when we take the modulus square of Eq.~\eqref{e_app_det_bs}.
The expectation value of Eq.~\eqref{e_app_n-eta} for an arbitrary input state in the other modes directly leads to
\begin{subequations}\label{e_app_Delta_N_eta}
\bg
N_{\eta \text{d}}= \eta_\text{d} N_\text{d} 
\eg
and
\bg
\Delta N_{\eta \text{d}}^2= \eta_\text{d}^2 \Delta N_\text{d}^2 + \eta_\text{d} (1-\eta_\text{d})N_\text{d},
\eg
\end{subequations}
where $N_{\eta \text{d}}$ and $\Delta N_{\eta \text{d}}^2$ are the photon number and variance detected by D$_\text{d}$, and $N_{\text{d}}$ and $\Delta N_{ \text{d}}^2$ the photon number and variance without detection loss.
With Eq.~\eqref{e_app_Delta_N_eta} we find for the phase sensitivity $\Delta \phi^2_{\eta\text{d}} \equiv \Delta N_{\eta \text{d}}^2 \Big/ \left| \frac{\partial N_{\eta \text{d}}}{\partial \phi}\right|^2$ including detection loss the expression
\bg\label{e_app_Delta_phi_eta}
\Delta \phi^2_{\eta\text{d}} = \Delta N_{\text{d}}^2 \Big/ \left| \frac{\partial N_{\text{d}}}{\partial \phi}\right|^2 \times \left( 1 + \frac{1-\eta_\text{d}}{\eta_\text{d}} \frac{N_\text{d}}{\Delta N^2_\text{d}} \right).
\eg

The above expressions are so far general for generic input in modes $\hat{a}$ and $\hat{\ell}$.
But now we make the assumption that we have a vacuum input in all modes.
When we rewrite Eq.~\eqref{e_app_a_prime_prime_prime} as $
\hat{b}\equiv A_\text{d} \hat{a} + \alpha_\text{d} \hat{a}^\dagger + B_\text{d} \hat{\ell} + \beta_\text{d} \hat{\ell}^\dagger 
$
and introduce the complex coefficients
\bg\label{e_coefficients_degenerate}
\begin{split}
A_\text{d} =  t_\text{d}u_\text{A} u_\text{B} +t_\text{d}^* v_\text{A}^* v_\text{B}\text{~,~}
B_\text{d} =   r_\text{d} u_\text{B}\text{~,~}\\
\alpha_\text{d} =    t_\text{d}v_\text{A} u_\text{B} +t_\text{d}^* u_\text{A}^* v_\text{B} \text{~and~}
\beta_\text{d} = r_\text{d}^* v_\text{B} , 
\end{split}
\eg
we see that $\hat{b} \ket{0}= \alpha_\text{d} \ket{1_{a}}+\beta_\text{d} \ket{1_{\ell}}$ and find for the state $\ket{\psi_\text{d}} \equiv\hat{b}^\dagger \hat{b} \ket{0}$ the expression
\bg
\begin{split}
\ket{\psi_\text{d}} = &\left( |\alpha_\text{d}|^2 + |\beta_\text{d}|^2 \right) \ket{0} + \sqrt{2} A_\text{d}^* \alpha_\text{d} \ket{2_{a}} \\
&+\left(B_\text{d}^* \alpha_\text{d}+A_\text{d}^* \beta_\text{d}\right) \ket{1_{a},1_{\ell}}+\sqrt{2}B_\text{d}^* \beta_\text{d} \ket{2_{\ell}}.
\end{split}
\eg
Hence, the vacuum expectation value $N_\text{d}\equiv \braket{0}{\psi_\text{d}}= |\alpha_\text{d}|^2 + |\beta_\text{d}|^2$ takes with Eq.~\eqref{e_coefficients_degenerate} the form
\bg
N_\text{d} = T_\text{d} V_\text{A} + V_\text{B} + 2 T_\text{d} V_\text{A} V_\text{B} - 2 T \sqrt{U_\text{A} V_\text{A} U_\text{B} V_\text{B}} \cos \phi,
\eg
where we used $R_\text{d}+T_\text{d}=1$, $U_j=1+V_j$, and introduced the phase
\bg \label{e_app_phase_deg}
\phi\equiv \operatorname{arg}\left(u_\text{A} v_\text{A} u_\text{B} v_\text{B}^* t_\text{d}^2\right)+\pi.
\eg
Note that the definition of the phase includes a shift by $\pi$ so that $\phi =0$ describes the dark fringe.
The variance can be calculated through $\Delta N_\text{d}^2 =\braket{\psi_\text{d}}{\psi_\text{d}}-N_\text{d}^2$ and we find after some algebra
\bg
\Delta N^2_\text{d} =  2N_\text{d} (1+ N_\text{d}) - R_\text{d} T_\text{d} V_\text{A}. 
\eg

\section{Quantum Fisher information}
\label{app_Quantum_Fisher_information}

In this appendix we calculate the quantum Fisher information for a degenerate NLI with vacuum input and equal gain in both crystals.
For a more convenient description, we use the Bogoliubov transformation from Eq.~\eqref{e_a_prime} with $u_\text{A}=u_\text{B}=u$ and $v_\text{A}=v_\text{B}=v$ to write the squeezed photon operator $\hat{a}' =\hat{S}^\dagger \hat{a}\hat{S}  $, where we introduced the squeezing operator $\hat{S}$.
If the gain is equal in both crystals and no loss is present, the final state at the output of the NLI is a pure state which can be written as a sequence of squeezing, phase evolution, and anti-squeezing.
Hence, it takes the form
\bg
\ket{\psi_f}= \hat{S}^\dagger \exp\left(\I \frac{\phi}{2}\hat{a}^\dagger \hat{a}\right)\hat{S}\ket{0}= \exp\left(\I \frac{\phi}{2}\hat{a}^{\prime\dagger} \hat{a}'\right)\ket{0}.
\eg
With the notation $\hat{n}' \equiv \hat{a}^{\dagger\prime} \hat{a}'$, the derivative of the final state with respect to $\phi$ can be written as $|\psi_f'\rangle = \hat{n}' \ket{\psi_f}/2$.

For a pure state, the quantum Fisher information~\cite{Pezze14} of the NLI can be written as
\bg
F_\phi=4( \braket{\psi_f'}{\psi_f'}-|\braket{\psi_f'}{\psi_f}|^2)= \left\langle \hat{n}'^2\right\rangle-\left\langle \hat{n}'\right\rangle ^2,
\eg
where the expectation values are taken with respect to the initial state, i.e., $\ket{0}$ in our case.
With the help of Eq.~\eqref{e_a_prime} we find the relation
\bg
\hat{n}'\ket{0}= V \ket{0}+ \sqrt{2} u^* v\ket{2},
\eg
and, by projecting this state on itself and on $\ket{0}$ we find the variance of $\hat{n}'$ and therefore show that the quantum Fisher information can be written as
\bg
F_\phi = 2 UV.
\eg

\section{Nondegenerate configuration}
\label{app_Non-degenterate_configuration}

In contrast to the degenerate case, we have for a nondegenerate setup two input modes, namely, modes 1 and 2, which are described by the bosonic annihilation operators $\hat{a}_1$ and $\hat{a}_2$, according to Fig.~\ref{fig_nondegenerate}.
Crystal A is described by the Bogoliubov transformation
\bg
\hat{a}_1' = u_\text{A} \hat{a}_1 + v_\text{A} \hat{a}_2^\dagger \quad\text{ and }\quad \hat{a}_2' = u_\text{A} \hat{a}_2 + v_\text{A} \hat{a}_1^\dagger.
\eg
We define the coefficients $v_\text{A}$ and $u_\text{A}$ in complete analogy to the degenerate case in Eq.~\eqref{e_a_prime}.

To model the loss that occurs inside the interferometer, we place two beam splitters in each branch, whose transmitted outputs is the input of crystal B.
We describe the beam splitter S$_j$ with $j= 1,2$ that accounts for internal loss through the transformation
\bg
 \hat{a}_j '' = t_j \hat{a}_j' + r_j \hat{\ell}_j \quad\text{and}\quad  \hat{\ell}_j ' = t_1^* \hat{\ell}_j -r_j^* \hat{a}_j'  .
\eg
Here, $r_j$ and $t_j$ denote the amplitude reflectivity and transmittance of the beam splitter.
In addition, we use the conventional definitions $R_j = |r_j|^2$ and $T_j = |t_j|^2$ as well as the relation $R_j+T_j = 1$.
The operators $\hat{\ell}_j$ describe the noise input of each beam splitter, the operators $\hat{\ell}_j'$ the output according to Fig.~\ref{fig_nondegenerate}.
The output of crystal B is then found through the relation
\bg
\hat{b}_1 = u_\text{B} \hat{a}_1'' + v_\text{B} \hat{a}_2^{\prime\prime \dagger} \quad\text{ and }\quad \hat{b}_2 = u_\text{B} \hat{a}_2'' + v_\text{B} \hat{a}_1^{\prime\prime\dagger}.
\eg
We define the coefficients $v_\text{B}$ and $u_\text{B}$ as in Eq.~\eqref{e_app_a_prime_prime_prime}.
Detection loss is modeled by the transformation
\bg
\hat{b}'_j= \sqrt{\eta_j} \hat{b}_j + \sqrt{1-\eta_j} \hat{d}_j,
\eg
where $\eta_j$ is the efficiency of the detector in output mode $j=1,2$ and $\hat{d}_j$ the noise that is introduced.
Since this transformation is completely analogous to Eq.~\eqref{e_app_det_loss}, we find exactly Eqs.~\eqref{e_app_Delta_N_eta} and~\eqref{e_app_Delta_phi_eta}, with the index $\text{d}$ now replaced by $j=1,2$.

With all the transformations above, including the beam splitters S$_{\eta_j}$ for detection loss, we find
\bg
\begin{split}
\hat{b}'_{1,2} = &A_{1,2} \hat{a}_{1,2} + \alpha_{1,2} \hat{a}_{2,1}^\dagger + B_{1,2} \hat{\ell}_{1,2}\\
&+ \beta_{1,2} \hat{\ell}_{2,1}^\dagger + \sqrt{1-\eta_{1,2}}\hat{d}_{1,2}
\end{split}
\eg
for the field detected by detector D$_{1,2}$.
Here, we defined the complex coefficients
\bg
\begin{split}
A_{1,2} =& \sqrt{\eta_{1,2}} (t_{1,2} u_\text{A}u_\text{B} + t_{2,1}^* v_\text{A}^* v_\text{B}),\\
B_{1,2} = &\sqrt{\eta_{1,2}}r_{1,2} u_\text{B},\\
\alpha_{1,2} = & \sqrt{\eta_{1,2}} (t_{1,2} v_\text{A}u_\text{B} + t_{2,1}^* u_\text{A}^* v_\text{B})  ~\text{ and }\\
\quad \beta_{1,2} =&\sqrt{\eta_{1,2}} r_{2,1}^* v_\text{B}.
\end{split}
\eg

It is straightforward to calculate $\hat{b}'_{1,2}\ket{0}= \alpha_{1,2} \ket{1_{a_{2,1}}}+\beta_{1,2} \ket{1_{\ell_{2,1}}}$.
With that result we find, in analogy to the calculation in the degenerate setup, for $\ket{\psi_j} \equiv\hat{b}^{\prime\dagger}_{1,2} \hat{b}'_{1,2}\ket{0}$ the expression
\bg\label{e_app_psi_12}
\begin{split}
\ket{\psi_{1,2}} = \left( |\alpha_{1,2}|^2 + |\beta_{1,2}|^2 \right) \ket{0} + \alpha_{1,2} A_{1,2}^* \ket{1_{a_1},1_{a_2}}\\
+\alpha_{1,2} B_{1,2}^* \ket{1_{a_{2,1}},1_{\ell_{1,2}}}+\beta_{1,2} A_{1,2}^* \ket{1_{a_{1,2}},1_{\ell_{2,1}}}\\
+\sqrt{1-\eta_{1,2}}\left(\alpha_{1,2}\ket{1_{a_{2,1}}}
+\beta_{1,2}\ket{1_{\ell_{2,1}}}\right)\ket{1_{d_{1,2}}}\\
+\beta_{1,2} B_{1,2}^* \ket{1_{\ell_1},1_{\ell_2}}.
\end{split}
\eg

It is easy to see that the photon number $N_{\eta j} = \braket{0}{\psi_j} = |\alpha_{j}|^2+|\beta_{j}|^2$ detected by D$_j$ takes the form
\bg\label{e_app_N_eta_j}
N_{\eta j}(\phi) = \eta_{j} \left( \mathcal{A}_j- \mathcal{K}_j\cos \phi\right)
\eg
with the amplitude $\mathcal{A}_j\equiv T_{j} V_\text{A} + V_\text{B} + 2 (T_1+T_2) V_\text{A} V_\text{B} $ and the contrast $\mathcal{K}_j=2 \sqrt{T_1 T_2 U_\text{A} U_\text{B} V_\text{A} V_\text{B}}$.
Note that the term in parentheses can be defined as the photon number $N_j$ without detection loss.
Moreover, the phase 
\bg
\phi \equiv \operatorname{arg}\left(u_\text{A} u_\text{B} v_\text{A} v_\text{B}^* t_{1}t_{2} \right)+\pi
\eg
is slightly differently defined from Eq.~\eqref{e_app_phase_deg} to include a phase that may be accumulated in the two arms in the interferometer and is included in the complex values of $t_j$.

When we calculate the variance $\Delta N^2_{\eta j}\equiv \braket{\psi_j}{\psi_j}-N_{\eta j}^2$ we find with the help of Eq.~\eqref{e_app_psi_12}
\bg
\Delta N_{\eta j}^2 = \left( |\alpha_{j}|^2+|\beta_{j}|^2\right) \left( 1-\eta_j + |A_{j}|^2+|B_{j}|^2 \right).
\eg
With the use of $ |A_{j}|^2+|B_{j}|^2= \eta_j+|\alpha_{j}|^2+|\beta_{j}|^2$, as well as Eq.~\eqref{e_app_N_eta_j}, this expression reduces to
\bg
\Delta N_{\eta j}^2 =  N_{\eta j} \left( 1 + N_{\eta j} \right).
\eg
It also implies directly that $\Delta N_j^2 = N_j (1+N_j)$ in the case without detection loss.
Moreover, the variance of the sum of both signals is
\bg
\Delta N_+^2 = \Delta N_{\eta 1}^2 + \Delta N_{\eta 2}^2 +  \braket{\psi_1}{\psi_2}+\braket{\psi_2}{\psi_1}-2N_{\eta 1} N_{\eta 2} .
\eg
The overlap
\bg
\braket{\psi_1}{\psi_2} =N_{\eta 1} N_{\eta 2}  + (\alpha_2 A_1 + \beta_2 B_1)(\alpha_1 A_2 + \beta_1 B_2)^*
\eg
takes a simple form.
Calculating the product is cumbersome, but using trigonometric relations, the definition of the phase $\phi$, the relations $U_j  = 1 + V_j$ as well as $T_j+R_j=1$, and Eq.~\eqref{e_app_N_eta_j} we arrive at
\bg
\begin{split}
\Delta N_+^2 =& (N_{\eta 1}+N_{\eta 2}) (1+N_{\eta 1}+N_{\eta 2})+ \eta_2 N_{\eta 1} + \eta_1 N_{\eta 2}\\&+\eta_1\eta_2  [2 T_1 T_2 - (T_1+T_2)]V_\text{A}.
\end{split}
\eg

\bibliography{bibliography}

\end{document}